\documentclass[epj,final]{svjour}
\usepackage[utf8]{inputenc}

\newcommand{\clover}[1]{$\mathrm{CLOVER}$#1}
\newcommand{\labr}[1]{LaBr$_3$:Ce#1}

\usepackage{graphicx}
\usepackage{hyperref} 

\usepackage{amssymb}
\usepackage{amsfonts}
\usepackage{amsmath}
\allowdisplaybreaks[4]

\usepackage[T1]{fontenc}
\usepackage[english]{babel}

\usepackage[
numbers, 
square, 
comma, 
sort&compress]{natbib} 
\title{First application of the Oslo Method in inverse kinematics}
\subtitle{Nuclear level densities and $\gamma$-ray strength functions of $^{87}\mathrm{Kr}$}

\author{
V. W. Ingeberg\inst{1} \and
S. Siem\inst{1} \and
M. Wiedeking\inst{2} \and
K.~Sieja\inst{3,4} \and
D.L.~Bleuel\inst{5} \and
C.P.~Brits\inst{2,6} \and
T.D.~Bucher\inst{2} \and
T.S.~Dinoko\inst{2} \and
J.L.~Easton\inst{2,7} \and
A.~Görgen \inst{1} \and
M.~Guttormsen\inst{1} \and
P.~Jones\inst{2} \and
B.V.~Kheswa\inst{2,8} \and
N.A.~Khumalo\inst{2} \and
A.C.~Larsen\inst{1} \and
E.A.~Lawrie\inst{2} \and
J.J.~Lawrie\inst{2} \and
S.N.T.~Majola\inst{2,9} \thanks{\emph{Present address:} Department of Physics, University of Johannesburg, P.O. Box 524, Auckland Park 2006, South Africa} \and
K.L.~Malatji\inst{2,6} \and
L.~Makhathini\inst{2,6} \and
B.~Maqabuka\inst{2,7} \and
D.~Negi\inst{2} \and
S.P.~Noncolela\inst{2,7} \and
P.~Papka\inst{2,6} \and
E.~Sahin\inst{1} \and
R.~Schwengner\inst{10} \and
G.M.~Tveten\inst{1} \and
F.~Zeiser\inst{1} \and
B.R.~Zikhali\inst{2,9}
}

\mail{vetlewi@fys.uio.no}

\institute{
Department of Physics, University of Oslo, N-0316 Oslo, Norway \and
iThemba LABS, P.O. Box 722, 7129 Somerset West, South Africa \and
Universit\'e de Strasbourg, IPHC, 23 rue du Loess 67037 Strasbourg, France \and
CNRS, UMR7178, 67037 Strasbourg, France \and
Lawrence Livermore National Laboratory, 7000 East Avenue, Livermore, California 94550-9234, USA \and
Department of Physics, Stellenbosch University, Private Bag X1, Matieland, 7602, South Africa \and
Department of Physics, University of the Western Cape, P/B X17 Bellville 7535, South Africa \and
Department of Physics, University of Johannesburg, P.O. Box 524, Auckland Park 2006, South Africa \and
Department of Physics, University of Zululand, Private Bag X1001, KwaDlangezwa 3886, South Africa \and
Institut f\"ur Strahlenphysik, Helmholtz-Zentrum Dresden-Rossendorf, 01328 Dresden, Germany
}

\date{\today}

\begin{document}

\abstract{The $\gamma$-ray strength function ($\gamma$SF) and nuclear level density (NLD) have been extracted for the first time from inverse kinematic reactions with the Oslo Method. This novel technique allows measurements of these properties across a wide range of previously inaccessible nuclei. Proton-$\gamma$ coincidence events from the  $\mathrm{d}(^{86}\mathrm{Kr}, \mathrm{p}\gamma)^{87}\mathrm{Kr}$ reaction were measured at iThemba LABS and the $\gamma$SF and NLD in $^{87}\mathrm{Kr}$ obtained. The low-energy region of the $\gamma$SF is compared to Shell Model calculations which suggest this region to be dominated by M1 strength. The $\gamma$SF and NLD are used as input parameters to Hauser-Feshbach calculations to constrain $(\mathrm{n},\gamma)$ cross sections of nuclei using the TALYS reaction code. These results are compared to $^{86}\mathrm{Kr}(n,\gamma)$ data from direct measurements.
}

\maketitle

\section{Introduction}\label{sec:Intro}
The nuclear level density (NLD) and the $\gamma$-ray strength function ($\gamma$SF) are fundamental properties of the nucleus. The NLD was introduced by Bethe soon after the composition of nuclei was firmly established \cite{PhysRev.50.332}. When excitation energy in a nucleus increases towards the particle separation energy, the NLD increases rapidly, creating a region referred to as the quasi-continuum. The ability of atomic nuclei to emit and absorb photons in the quasi-continuum is determined by the $\gamma$SF \cite{Bart1973}. It is a measure of the average reduced $\gamma$-ray decay probability and reveals essential information about the electromagnetic response and therefore the nuclear structure of the nucleus.

With their significant applicability to astrophysical element formation via capture processes \cite{0034-4885-76-6-066201,ARNOULD20031, ARNOULD200797,GORIELY199810}, NLDs and $\gamma$SFs have received increased experimental and theoretical attention \cite{Larsen2019}. They are also relevant to the design of existing and future nuclear power reactors, where reactor simulations depend on many evaluated nuclear reactions \cite{ENDF_BVII.1,Chadwick2011}. The importance of NLDs and $\gamma$SFs is increasingly being recognized and a reference database for $\gamma$SFs has been established \cite{Goriely2019}.
Nonetheless, challenges remain and nuclear physics properties, such as the NLD and $\gamma$SF, remain a main source of uncertainty in cross-section calculations. This is either due to the complete lack of experimental data or the associated large experimental uncertainties. 

The situation can be improved through accurate experimental neutron capture cross sections, or indirectly by measuring NLD and $\gamma$SF data. One experimental approach, the Oslo Method \cite{OsloMethodNIM}, has been extensively used to measure the NLD and $\gamma$SF from particle-$\gamma$ coincident data. 
NLDs and $\gamma$SFs obtained with the Oslo Method have been shown to provide reliable neutron capture cross sections \cite{PhysRevC.96.024313,PhysRevC.95.045805} and proton capture cross sections \cite{PhysRevC.93.045810}. In recent years, the Oslo Method has been extended to extract the $\gamma$SF and NLD following $\beta$ decay \cite{PhysRevLett.113.232502}. 
Using $\gamma$SFs and NLDs to determine capture cross sections has several advantages since these properties can be obtained for any nucleus that can be  populated in a reaction from which the excitation energy can be experimentally determined. Although the Oslo and $\beta$-Oslo Methods provide access to a vast range of stable and radioactive nuclei some species remain inaccessible. Many more nuclei become accessible by using inverse kinematic reactions, from radioactive species to several stable isotopes for which the manufacture of targets is problematic due to their chemical or physical properties. 

In this Letter we report on the first application to measure the NLD and $\gamma$SF with the Oslo Method following an inverse kinematic reaction. This work lays the foundation of new opportunities to study statistical properties of nuclei, which were previously inaccessible, at stable and radioactive ion beam facilities. 
The results from the $\mathrm{d}(^{86}\mathrm{Kr},\mathrm{p})^{87}\mathrm{Kr}$ reaction exhibit a low-energy enhancement of the $\gamma$SF in $^{87}\mathrm{Kr}$, which is discussed in the context of Shell Model calculations. The $^{86}\mathrm{Kr}(\mathrm{n},\gamma)$ cross section is obtained from the TALYS reaction code \cite{Koning2007} and compared to previous direct measurements to test the robustness of the experimental method. 

\section{Experiment}\label{sec:Exp}
The experiment was performed with a $300$~MeV $^{86}\mathrm{Kr}$ beam from the Separated Sector Cyclotron facility at iThemba LABS. Polyethylene targets with $99\%$ deuteron enrichment were bombarded with a beam intensity of $\approx 0.1$ pnA for $80$ hrs. Several deuterated polyethylene targets, ranging in thicknesses from $110$ to $550$ $\mathrm{\mu}$g/cm$^2$, were used.
Accounting for the target thicknesses the center-of-mass (CM) energy was $6.44(40)$ MeV. The reactions were identified through the detection of light charged particles in two silicon $\mathrm{\Delta}$E-E telescopes covering scattering angles between $24^\circ$ and $67^\circ$ relative to the beam direction (corresponding to CM-angles $38^\circ$ to $121^\circ$). The E detectors were 1 mm thick while the $\mathrm{\Delta}$E detectors were 0.3 and 0.5 mm thick. The dimensions of the W1-type double-sided silicon strip detectors \cite{micron} were $4.8 \times 4.8$ cm and they consisted of 16 parallel and perpendicular strips $3$ mm wide with an opening angle of $\approx 1.5^\circ$ for each pixel. Suppression of $\delta$ electrons was achieved by an aluminum foil of $4.1$ mg/cm$^2$ areal density which was placed in front of the $\mathrm{\Delta}$E detectors.
The $\gamma$-rays were measured with the AFRODITE array \cite{NIMA2007}, which at the time of the experiment consisted of eight collimated and Compton suppressed high-purity germanium \clover{}-type detectors. Two non-collimated \labr{} detectors ($3.5^"\times8^"$) were coupled to the AFRODITE array and mounted $24$ cm from the target at $45^\circ$. The detectors were calibrated using standard $^{152}\mathrm{Eu}$ and $^{56}\mathrm{Co}$ sources. 
The detector signals were processed by XIA digital electronics in time-stamped list mode with each channel self-triggered.

From the time-stamped list mode data, entries were selected based on their time-stamps being within a window of $\pm1850$ ns in an E-detector entry. The ratio of energy deposited in the $\Delta$E- to the E-detector is used to determine the outgoing reaction channels. The selection of proton-$\gamma$ events was made with an 80 ns wide time-gate on the prompt time peak. Contributions from uncorrelated events were subtracted from the data by placing off-prompt time gates of equal length. This leads to   approximately 100k proton-$\gamma$ events in both \labr{} and \clover{} matrices.
In this letter only the data from the \labr{} detectors are included, although data from the \clover{} detectors yield similar results.
Kinematic corrections due to the reaction Q-value, recoil energy of $^{87}\mathrm{Kr}$, and the energy losses of the protons in the target and aluminum foils were applied to determine the excitation energy of the populated states, with the a resulting FWHM for excitation energy of $\approx 1$ MeV. The $\gamma$-rays in coincidence with protons were Doppler corrected by assuming the residual $^{87}\mathrm{Kr}$ nucleus not being deflected from the beam axis and has a constant velocity of $8.5\%$ of c. Due to these assumptions the error in deflection angle is less than $1.3^\circ$ while the error in velocity is less than $0.4\%$ of c. These error are negligible as the major contributor to errors in the Doppler correction is the $17^\circ$ opening angle of the LaBr3:Ce detectors. 
Background from $^{86}\mathrm{Kr}+^{12}\mathrm{C}$ fusion evaporation events have been simulated with PACE4 \cite{LisePP} and found to have a very low proton yield ($< 4\%$) with proton energies outside the energy range considered in the analysis.
This matrix is unfolded \cite{UnfoldingNIM} with response functions of the detectors extracted from a Geant4 \cite{AGOSTINELLI2003250} simulation of the \labr{} detectors.
An iterative subtraction method, known as the First-Generation Method \cite{FirstGenerationNIM}, is applied to the unfolded $\gamma$-ray spectra, revealing the distribution of primary $\gamma$-rays in each excitation bin (256 keV bin width for both the $E_x$ and $E_\gamma$ axes).

The NLD $\rho(E_x)$ at excitation energy $E_x$ and $\gamma$-ray transmission coefficient, $\mathcal{T}(E_\gamma)$, are related to the primary $\gamma$-ray spectrum by \cite{OsloMethodNIM}
\begin{equation}
 P(E_x, E_\gamma) \propto \rho(E_x - E_\gamma) \mathcal{T}(E_\gamma), 
 \label{eq:PrimaryMat}
\end{equation}
and are extracted with a $\chi^2$-method \cite{OsloMethodNIM} giving the unique solution of the functional shape of the NLD and $\mathcal{T}(E_\gamma)$. These are normalized to known experimental data to retrieve the correct slope and absolute value. The extraction has been performed within the limits $3.2<E_x<5.2$ MeV and $E_\gamma>1.7$ MeV of the primary $\gamma$-ray matrix where the level density is sufficiently high for statistical decay to be dominant.

\section{Normalization}\label{sec:Normalization}
From the primary $\gamma$-ray spectrum the NLD $\widetilde{\rho}(E_x)$ and $\gamma$-transmission coefficient $\widetilde{\mathcal{T}}(E_\gamma)$ are extracted. These are related to the physical solution by the following transformation \cite{OsloMethodNIM}:
\begin{align}
    \rho(E_x) &= A\widetilde{\rho}(E_x)e^{\alpha E_x} \label{eq:primary_transform_R} \\
    \mathcal{T}(E_\gamma) &= B\widetilde{\mathcal{T}}(E_\gamma)e^{\alpha E_\gamma} \label{eq:primary_transform_T},
\end{align}
where A and B are the absolute values for the level density and the transmission coefficient, respectively, and $\alpha$ is the common slope parameter.

For the level density, the slope and absolute value are determined by a fit to the level density found from the known discrete levels \cite{JOHNSON20151} at low-excitation energy and the level density at the neutron separation energy ($S_n = 5.5$ MeV). The level density of $J=1/2$ levels at $S_n$ is determined from the average resonance spacing of s-wave resonances ($D_0$) and p-wave resonances ($D_1(J=1/2)$) by
\begin{equation}
\rho(Sn,J=1/2) = \frac{1}{D_0} + \frac{1}{D_1(J=1/2)}.\label{eq:rhoSnOneHalf}
\end{equation}
Where the spacing parameters are taken from \cite{PhysRevC.38.1605}. The full level density at $S_n$ is determined by
\begin{equation}
\rho(S_n) = \rho(S_n,J=1/2)/g(Sn,J=1/2), \label{eq:FullRhoSn}
\end{equation}
where g is the spin distribution \cite{Ericson1958}
\begin{equation}
    g(E, J) = \frac{2J + 1}{2\sigma^2(E)} e^{-(J + 1/2)^2/2\sigma^2(E)}. \label{eq:SpinDistribution}
\end{equation}
The spin cutoff parameter $\sigma(E)$ is modeled with the following energy dependence \cite{PhysRevC.96.024313} 
\begin{equation}
    \sigma^2(E) = \sigma_d^2 + \frac{E - E_d}{S_n - E_d}(\sigma^2(S_n) - \sigma_d^2), \label{eq:SpinCutShape}
\end{equation}
\noindent where $E_d$ is the excitation energy below which the spin cutoff parameter $\sigma = \sigma_d$ is a constant. The spin cutoff parameter $\sigma_d$ at $E_d \leq 2.4$ MeV is estimated to be $1.75(26)$, based on the spin assignment of the known levels, while the cutoff parameter at the neutron separation energy $\sigma(S_n)$ is estimated to be $3.95(60)$, based on the predictions of the spin cutoff models of Refs. \cite{PhysRevC.72.044311,CanJPhys43.1446,PhysRevC.80.054310}. The shape of the spin distribution predicted by the Hartree-Fock-Bogoliubov plus combinatorial model \cite{PhysRevC.78.064307} has also been considered and found to be in agreement. Based on the estimated uncertainties of $\sigma(E)$ and the experimental uncertainties of the resonance spacing, the total NLD at $S_n$ is found to be $1472(427)$ MeV$^{-1}$.

The level density extracted with the Oslo Method extracted extends up to $3.7$ MeV and an interpolation between the Oslo Method data and the neutron separation energy has to be done. This interpolation uses the constant temperature (CT) shape \cite{doi:10.1080/00018736000101239}
\begin{equation}
\rho_\text{CT}(E) = \frac{1}{T}e^{\frac{E - E_0}{T}}, \label{eq:ConstT}
\end{equation}
with shift parameter $E_0 = S_n - T\ln(T\rho(S_n))$ to ensure that the interpolation matches the experimental known $\rho(S_n)$. The optimum temperature parameter $T$ in the interpolation, as well as the normalization parameters $A$ and $\alpha$, are determined through a least-squares fit between the level density extracted in the Oslo Method and the discrete levels for energies below $E_x = 2.4$ MeV and the CT interpolation above.

Since the reaction is sub-Coulomb barrier the primary reaction channel will be neutron capture following inelastic deuteron breakup in the Coulomb field of the $^{86}\mathrm{Kr}$ projectile and $1/2$ states are assumed to be strongly favored in the initial population, and has to be accounted for. Since the resulting normalized level density found with the Oslo Method will correspond to the level density of $1/2$ and $3/2$ levels, the total level density is recovered by dividing by $g(E_x,1/2)+g(E_x,3/2)$.

\begin{figure}
    \includegraphics[width=8.6cm]{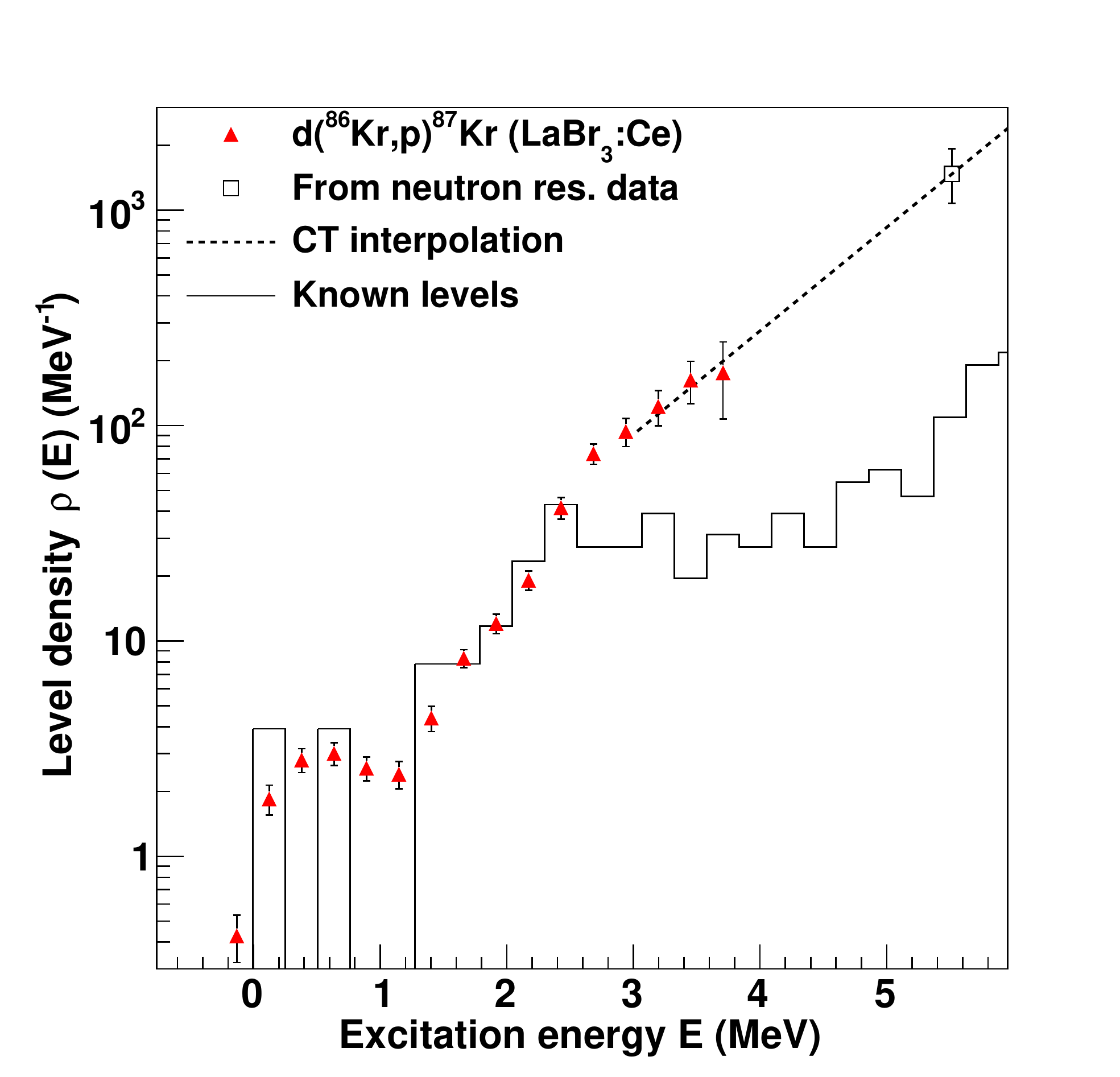}
    \caption{\label{fig:NLD}Normalized $^{87}\mathrm{Kr}$ nuclear level densities for \labr{} (red circles) detectors. The black line shows the known levels while the open square is the level density at the neutron separation energy. The dashed line is the constant temperature interpolation. The error bars represent the upper and lower uncertainty limit due to all known statistical and systematic effects.}
\end{figure}

The same normalization procedure has been repeated, but with an interpolation with a shape matching that of the Back-Shifted Fermi-gas model \cite{Erba1961,CanJPhys43.1446} with the difference in the resulting normalization included in the error-bars. All errors due to systematical and statistical effects of the Oslo Method \cite{PhysRevC.83.034315}, together with those related to the normalization process have been propagated to give the level density with error-bars shown in \autoref{fig:NLD}.

\begin{table}
    \centering
    \caption{\label{tab:normPar}Experimental values and parameters used in the normalization. The spin cutoff at $S_n$ $\sigma(Sn)$ is an average of the models presented in \cite{PhysRevC.72.044311,CanJPhys43.1446,PhysRevC.80.054310,PhysRevC.78.064307} while $\sigma_d$ is estimated from discrete states with known spin. The level density of $1/2$ levels at $S_n$ are found using eq. \eqref{eq:rhoSnOneHalf} and the total level density at $S_n$ with eq. \eqref{eq:FullRhoSn}. The temperature $T$ is determined from a least-squares fit to data points in the range $2.4 < E_x < 3.7$ MeV.}
            \begin{tabular}{l|c} \hline
            $D_0$ & $26.2(21)$ keV \cite{PhysRevC.38.1605} \\ \hline
            $D_1$ (J=1/2) & $18.8(14)$ keV \cite{PhysRevC.38.1605} \\ \hline
            $\sigma(S_n)$ & $3.95(60)$ \\ \hline
            $\sigma_d$ & $1.75(26)$ \\ \hline
            $\rho(S_n,1/2)$ & $91(5)$ MeV$^{-1}$ \\ \hline
            $\rho(S_n)$ & $1472(427)$ MeV$^{-1}$ \\ \hline
            $\langle \Gamma_{\gamma0} \rangle$ & $0.25(10)$ eV \cite{PhysRevC.28.602} \\ \hline
            $T$ & $0.9(1)$ MeV \\ \hline
            \end{tabular}
\end{table}

\begin{figure}
    \includegraphics[width=8.6cm]{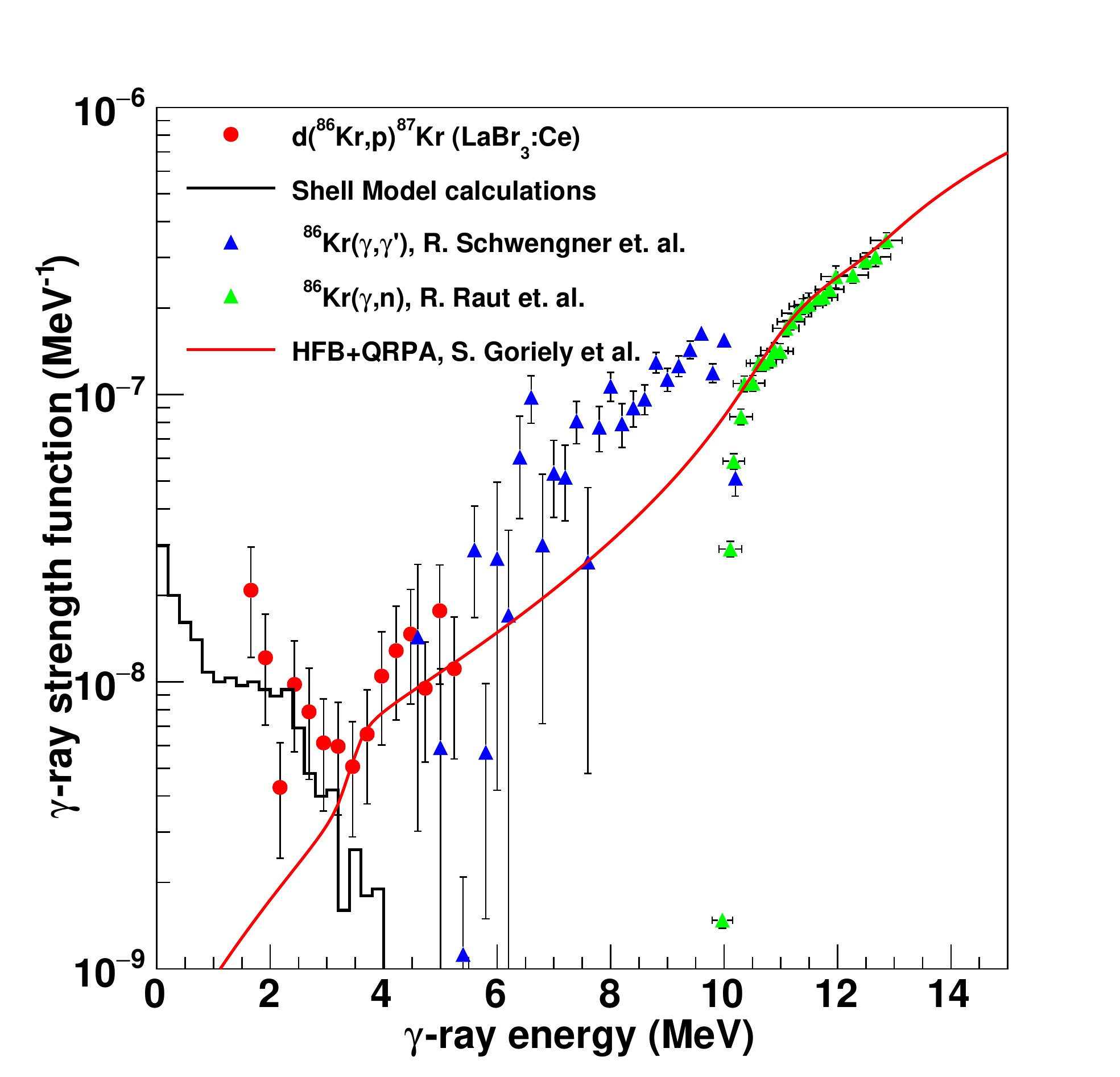}
    \caption{\label{fig:GSF} $\gamma$-ray strength function of $^{87}\mathrm{Kr}$ (red circles) compared with the $\gamma$-ray strength function of $^{86}\mathrm{Kr}$ extracted from $^{86}\mathrm{Kr}(\gamma,\gamma^\prime)$ (blue triangles) \cite{PhysRevC.87.024306} and $^{86}\mathrm{Kr}(\gamma,\mathrm{n})$ (green squares) \cite{PhysRevLett.111.112501}. The solid black line are results from Shell Model calculations with a $^{78}\mathrm{Ni}$ core (see \autoref{sec:SMC} for details), while the red line is the microscopic HFB+QRPA prediction \cite{GORIELY2004331} for the E1 strength. The error bars include all known statistical and systematic errors.}
\end{figure}

The absolute value of the transmission coefficients are normalized to the average radiative width of s-wave resonances $\langle \Gamma_{\gamma0} \rangle$ in a process detailed in \cite{PhysRevC.63.044313}, and converted to $\gamma$SF by $f(E_\gamma) = \mathcal{T}(E_\gamma)/(2\pi E_\gamma^3)$. The value of $\langle \Gamma_{\gamma0} \rangle$ is estimated to be $0.25(10)$ eV based on the measured $\Gamma_\gamma$ of s-wave resonances of \cite{PhysRevC.28.602}. The resulting $\gamma$SF with all errors propagated are shown in \autoref{fig:GSF}. All experimental values and parameters used in the normalization process are listed in \autoref{tab:normPar}.


\section{Nuclear level densities and $\gamma$-ray strength functions}\label{sec:NLD_gSF}

The normalized NLD is shown in \autoref{fig:NLD} and is in excellent agreement with the constant temperature level density and matches well with the known discrete states at lower excitation energies. 
The normalized $\gamma$SF is shown in \autoref{fig:GSF} and is consistent with $\gamma$SFs from $^{86}\mathrm{Kr}(\gamma,\gamma^\prime)$ \cite{PhysRevC.87.024306} and $^{86}\mathrm{Kr}(\gamma,\mathrm{n})$ \cite{PhysRevLett.111.112501}, with the enhancement seen in the $(\gamma,\gamma^\prime)$ data between $6$ and $8$ MeV caused by a Pygmy resonance \cite{PhysRevC.87.024306}.  A drop in the $\gamma$SFs at $\sim 2.1$ MeV is caused by the $2123$-keV state in $^{87}\mathrm{Kr}$, which is strongly populated in the reaction, but less through feeding from the quasi-continuum. This causes the first generation method to over-subtract in the higher excitation-energy bins, causing an artificial drop in the $\gamma$SF. This effect has previously been discussed \cite{PhysRevC.83.034315}. At low energies we observe a large enhancement in the $\gamma$SF, similar to what has been observed in several other nuclei  \cite{PhysRevC.93.064302,PhysRevC.71.044307,PhysRevC.73.064301,PhysRevC.80.044309,PhysRevC.87.014319,PhysRevC.93.034303,Kheswa2015268}. Although the \textit{upbend} has been independently confirmed \cite{PhysRevLett.108.162503}, little is known of the origin of this feature, except that it is dominated by dipole radiation \cite{PhysRevLett.111.242504,Lar2017,PhysRevC.97.024327} and that it can have large effects on neutron capture cross sections \cite{PhysRevC.82.014318}.

\section{Shell Model calculations}\label{sec:SMC}
Calculations within the shell-model framework predicts the upbend due to M1 transitions \cite{PhysRevLett.111.232504}. In this work, large-scale shell-model calculations of the M1 component of the $\gamma$SF were performed in the model space outside the $^{78}$Ni core, containing $f_{5/2}p_{3/2}p_{1/2}g_{9/2}$-proton and $d_{5/2}s_{1/2}\allowbreak d_{3/2}g_{7/2}h_{11/2}$-neutron orbitals. 
The effective interaction employed here is described e.g. in Refs. \cite{PhysRevC.92.014328,PhysRevC.92.064322}.  
The diagonalization of the Hamiltonian matrix in the full configuration space 
was achieved using the Strasbourg shell-model code NATHAN \cite{RevModPhys.77.427}. 
The spin-part of the magnetic operator was quenched by a common factor of 0.75 \cite{RevModPhys.77.427}.
We computed this way up to 60 states of each spin between $1/2$ and $15/2$ for both parities. 
This leads to a total of around $8\cdot10^4$ M1 matrix elements, among which 
$14822$ connect states located in the energy range $E_{x}=3.4-5.4$ MeV, as  considered in the experiment. To obtain the average strength per energy interval, $\langle B(M1)\rangle$, the total transition strength was accumulated in $200$ keV bins and divided by the number of transitions within these bins. 
The $\gamma$SF was obtained from the relation $f_{M1}(E_\gamma, E_i, J_i, \pi) = 16\pi/9(\hbar c)^{-3}\langle B(M1)(E_{\gamma}, E_i, J_i, \pi)\rangle\rho(E_i, J_i, \pi)$, where \linebreak $\rho_i (E_i, J_i, \pi)$ is the partial level density at the energy of the initial state ($E_i$). 
The $\gamma$SF, shown in \autoref{fig:GSF}, is an average of the $f_{M1}$s evaluated for each spin/parity separately.
The shape of the shell-model $\gamma$SF is consistent with experimental data up to $\sim3$MeV. Since 
the model space does not contain all spin-orbit partners (i.e., $\nu g_{9/2}$ and $\pi f_{7/2}$ orbits) the strength above 4 MeV, due to the spin-flip transitions, cannot be accounted for. However, the
theoretical $\gamma$SF exhibits significant strength at $E_\gamma=0$, as   
in the previous shell-model calculations in this mass region \cite{PhysRevLett.111.232504}.
The largest $B(M1)$ contributions at low $\gamma$-ray energies in $^{87}\mathrm{Kr}$ are related to transitions between close-lying negative-parity
states with $\nu d_{5/2}\otimes\pi f_{5/2}^{-1}g_{9/2}^1$ and $\nu d_{5/2}\otimes\pi p_{3/2}^{-1}g_{9/2}^1$ components. 
The magnitude of the theoretical M1 strength is in good agreement with the data as measured in the experiment, however we cannot exclude an additional contribution from E1 strength. Recent experimental results in $^{56}\mathrm{Fe}$ \cite{PhysRevC.97.024327} could suggest a mixture of M1 and E1 radiation in the enhancement region and the addition of a non-zero E1 component without an upbend towards $E_\gamma \rightarrow 0$ MeV 
is predicted from Shell Model calculations \cite{PhysRevLett.119.052502}. Including the E1 strength calculations from the Hartree-Fock-Bogolyubov + QRPA (HFB+QRPA) model by \cite{GORIELY2004331} we observe an overall good agreement between theoretical predictions and experimental results.

\section{Neutron capture cross sections}\label{sec:NCCS}
In a statistical framework the $^{86}\mathrm{Kr}(\mathrm{n},\gamma)$ cross section can be determined from the NLD, $\gamma$SF and a suitable neutron optical model potential (nOMP) for $^{87}\mathrm{Kr}$. Phenomenological nOMPs e.g. from Ref. \cite{Koning2003} are observed to give good agreement with the total cross section for nuclei close to the valley of stability. We performed Hauser-Feshbach (HF) \cite{PhysRev.87.366} calculations with the TALYS\footnote{Version 1.9} code \cite{Koning2007}, and the optical model potential of Ref. \cite{Koning2003}. A semi-microscopic optical model \cite{PhysRevC.63.024607} was also tested, and gave virtually the same results. Pre-equilibrium reactions were also taken into account. 
The $^{87}\mathrm{Kr}$ states used for the TALYS calculations are described by the known discrete states up to $2.3$ MeV, and by the measured NLD above. Beyond $3.7$ MeV the NLD is described by the CT interpolation.

\begin{figure}
    \includegraphics[width=8.6cm]{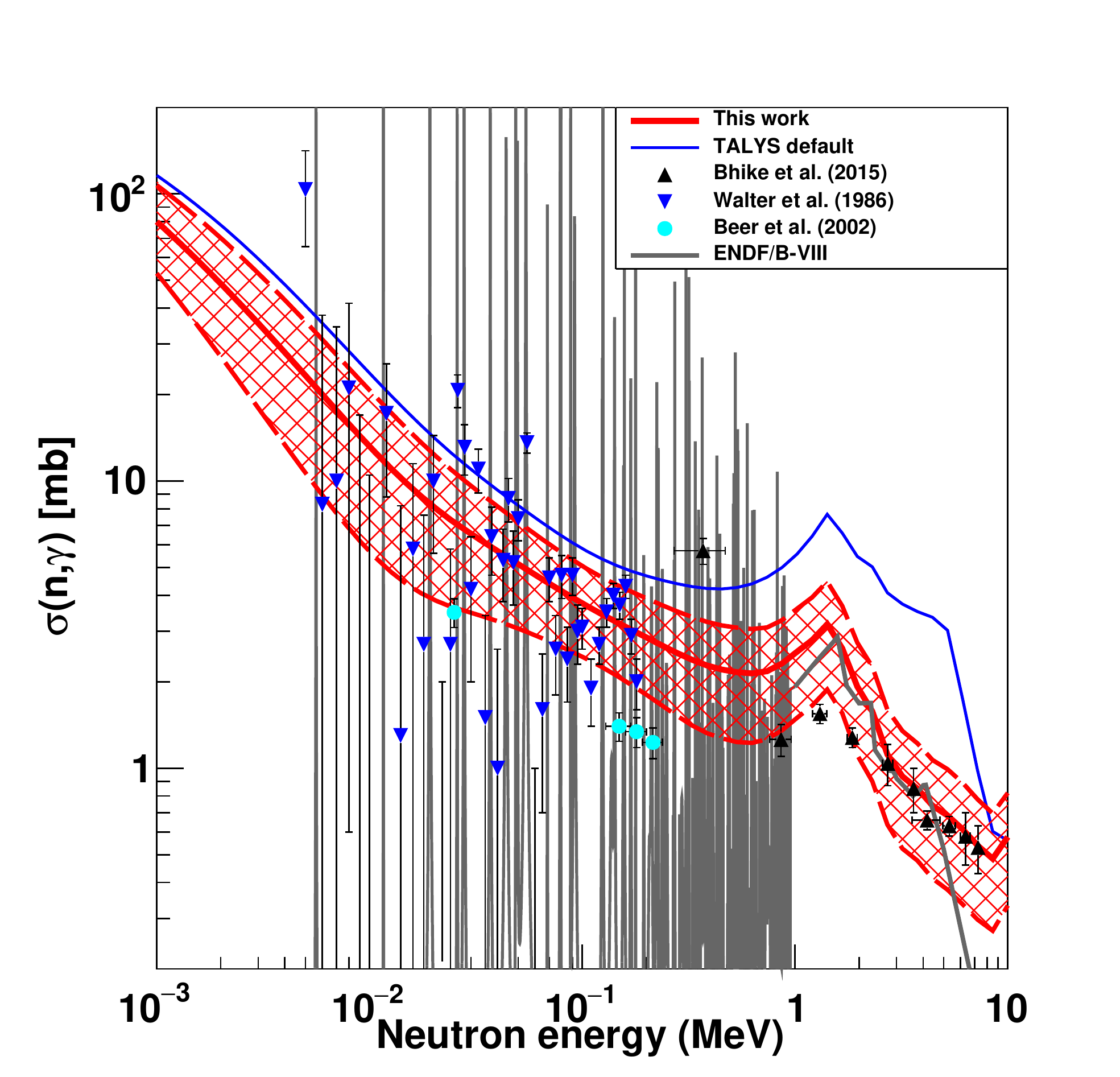}
    \caption{\label{fig:cs} $^{86}\mathrm{Kr}(n,\gamma)$ cross sections. The red-hashed area represents the total uncertainty based on both systematical and statistic errors. The gray and blue lines are from the evaluation of ENDF/B-VII.1 \cite{ENDF_BVII.1} and the TALYS default input, respectively, and is provided for comparison. The black triangles shows the direct measurements of Bhike \textit{et al.} \cite{PhysRevC.92.014624}, the blue upside-down triangles are results from time-of-flight measurements of Walter \textit{et al.} \cite{NuclSciEng93.357} and the turquoise circles are the results from the activation measurements of Beer \textit{et al.} \cite{Beer2002}.}
\end{figure}

The measured $\gamma$SF are used as input between $1.6 \leq E_\gamma \leq 5.2$ MeV (excluding the 2.1 MeV data point), as shown in \autoref{fig:GSF}, for $E_\gamma < 1.6$ MeV the results from the Shell Model calculations are used, while the results from microscopic HFB+QRPA calculations \cite{GORIELY2004331}, as implemented in TALYS, of the E1 strength are used for $E_\gamma > 5.2$ MeV. The M1 spin-flip contribution is also included as a standard Lorenzian with the TALYS parameterization.
\autoref{fig:cs} shows the resulting neutron capture cross section  calculations. The input parameters have been varied in accordance with the statistical and systematic uncertainties to produce the red hashed error-band. We observe an overall good agreement with direct measurements by Bhike \textit{et al.} \cite{PhysRevC.92.014624} and a decent agreement at higher energies with measurements of Walter \textit{et al.} \cite{NuclSciEng93.357}, while somewhat high compared with the activation results of Beer \textit{et al.} \cite{Beer2002}. The Maxwellian average (MACS) at the typical s-process temperature of 30 keV is found to be 7.6(49) mb, which is higher than the evaluated value of 3.4(3) mb found in KaDoNis \cite{Dillmann2006}. This discrepancy can be explained by the fact that HF calculations will give results that overestimate the MACS for low temperatures when the level density is low \cite{PhysRevC.56.1613}. A possible resolution could be to use Monte Carlo simulations to generate statistical resonances from average nuclear properties as proposed in \cite{Rochman2013,Rochman2017}.

\section{Conclusion}\label{sec:Conclution}
We have presented a novel method for obtaining $\gamma$SF and NLD using inverse kinematic reactions, which opens opportunities to study a wide range of stable and radioactive nuclei.
The $d(^{86}\mathrm{Kr},p\gamma)$ reaction was used to measure the NLD and $\gamma$SF in $^{87}\mathrm{Kr}$. The low-energy part of the $\gamma$SF is found to exhibit an enhancement. Shell Model calculations were performed and suggest that the enhancement is predominantly due to low-energy M1 transitions in $^{87}\mathrm{Kr}$.

The $\gamma$SF and NLD measurements in $^{87}\mathrm{Kr}$ were used to calculate $(n,\gamma)$ cross sections, which are in good agreement with those from direct measurements, and give confidence in the approach using inverse kinematic reactions. This is consistent with the findings of previous work with the Oslo Method and is particularly interesting since direct measurement of neutron capture cross sections over a wide range of incident neutron energies is very challenging. It is clear that  $\gamma$SFs and NLDs provide a viable alternative to obtain reliable capture cross sections. 
 
With inverse kinematics, new regions of the nuclear chart become accessible to experiments, which also brings about new challenges. For exotic nuclei, neutron resonance data are not known and the normalizing procedure needs to be revised. One possibility is that the slope of the $\gamma$SF, and thereby also the slope of the NLD, could be constrained using a technique where the ratio of populated discrete states from the quasi-continuum is used to determine the shape of the $\gamma$SF \cite{PhysRevLett.108.162503,PhysRevC.93.054311}, leaving the absolute value of the NLD to be determined by the known discrete levels. Unfortunately, this still does not determine the absolute value of the $\gamma$SF. However, reasonable estimates of the absolute value may be obtained from systematics of the $\langle \Gamma_{\gamma0}\rangle$. 

Measuring statistical properties of nuclei from inverse kinematic reactions provides a novel and complementary foundation for exploring the limitations of the current models of statistical behavior in the nucleus. It will allow to further constrain the uncertainties in models which are used in nuclear astrophysics and reactor physics. 

\begin{acknowledgement}
The authors would like to thank iThemba LABS operations for stable running conditions and John Greene (Argonne National Lab.) for providing excellent targets. This work is based on research supported by the Research Council of Norway under project Grants No. 222287, 262952 (G.M.T), 263030 (V.W.I, S.S, A.G, F.Z) and 240104 (E.S), by the National Research Foundation of South Africa under grant no 118846, and the U.S. Department of Energy by Lawrence Livermore National Laboratory under Contract DE-AC52-07NA27344. A.C.L. gratefully acknowledges funding through ERC-STG-2014 Grant Agreement No. 637686, and support from the ChETEC Cost Action (CA16117) supported by COST. This work was performed within the IAEA CRP on ``Updating the Photonuclear data Library and generating a Reference Database for Photon Strength Functions'' (F410 32). M. W. and S. S. acknowledge the support from the IAEA under Research Contract 20454 and 20447, respectively.
\end{acknowledgement}

\bibliography{biblography.bib}

\begin{thebibliography}{65}
\providecommand{\natexlab}[1]{#1}
\providecommand{\url}[1]{\texttt{#1}}
\expandafter\ifx\csname urlstyle\endcsname\relax
  \providecommand{\doi}[1]{doi: #1}\else
  \providecommand{\doi}{doi: \begingroup \urlstyle{rm}\Url}\fi

\bibitem[Bethe(1936)]{PhysRev.50.332}
H.~A. Bethe.
\newblock An attempt to calculate the number of energy levels of a heavy
  nucleus.
\newblock \emph{Phys. Rev.}, 50:\penalty0 332--341, Aug 1936.
\newblock \doi{10.1103/PhysRev.50.332}.
\newblock URL \url{https://link.aps.org/doi/10.1103/PhysRev.50.332}.

\bibitem[Bartholomew et~al.(1973)Bartholomew, Earle, Ferguson, Knowles, and
  Lone]{Bart1973}
G.~A. Bartholomew, E.~D. Earle, A.~J. Ferguson, J.~W. Knowles, and M.~A. Lone.
\newblock \emph{Gamma-Ray Strength Functions}, pages 229--324.
\newblock Springer US, Boston, MA, 1973.
\newblock ISBN 978-1-4615-9044-6.
\newblock \doi{10.1007/978-1-4615-9044-6_4}.
\newblock URL \url{http://dx.doi.org/10.1007/978-1-4615-9044-6_4}.

\bibitem[Rauscher et~al.(2013)Rauscher, Dauphas, Dillmann, Fröhlich, Fülöp,
  and Gyürky]{0034-4885-76-6-066201}
T~Rauscher, N~Dauphas, I~Dillmann, C~Fröhlich, Zs~Fülöp, and Gy~Gyürky.
\newblock Constraining the astrophysical origin of the p-nuclei through nuclear
  physics and meteoritic data.
\newblock \emph{Reports on Progress in Physics}, 76\penalty0 (6):\penalty0
  066201, 2013.
\newblock URL \url{http://stacks.iop.org/0034-4885/76/i=6/a=066201}.

\bibitem[Arnould and Goriely(2003)]{ARNOULD20031}
M.~Arnould and S.~Goriely.
\newblock The p-process of stellar nucleosynthesis: astrophysics and nuclear
  physics status.
\newblock \emph{Physics Reports}, 384\penalty0 (1):\penalty0 1 -- 84, 2003.
\newblock ISSN 0370-1573.
\newblock \doi{https://doi.org/10.1016/S0370-1573(03)00242-4}.
\newblock URL
  \url{http://www.sciencedirect.com/science/article/pii/S0370157303002424}.

\bibitem[Arnould et~al.(2007)Arnould, Goriely, and Takahashi]{ARNOULD200797}
M.~Arnould, S.~Goriely, and K.~Takahashi.
\newblock The r-process of stellar nucleosynthesis: Astrophysics and nuclear
  physics achievements and mysteries.
\newblock \emph{Physics Reports}, 450\penalty0 (4):\penalty0 97 -- 213, 2007.
\newblock ISSN 0370-1573.
\newblock \doi{https://doi.org/10.1016/j.physrep.2007.06.002}.
\newblock URL
  \url{http://www.sciencedirect.com/science/article/pii/S0370157307002438}.

\bibitem[Goriely(1998)]{GORIELY199810}
S.~Goriely.
\newblock Radiative neutron captures by neutron-rich nuclei and the r-process
  nucleosynthesis.
\newblock \emph{Physics Letters B}, 436\penalty0 (1):\penalty0 10 -- 18, 1998.
\newblock ISSN 0370-2693.
\newblock \doi{https://doi.org/10.1016/S0370-2693(98)00907-1}.
\newblock URL
  \url{http://www.sciencedirect.com/science/article/pii/S0370269398009071}.

\bibitem[Larsen et~al.(2019)Larsen, Spyrou, Liddick, and
  Guttormsen]{Larsen2019}
A.~C. Larsen, A.~Spyrou, S.~N. Liddick, and M.~Guttormsen.
\newblock {Novel techniques for constraining neutron-capture rates relevant for
  r-process heavy-element nucleosynthesis}, apr 2019.
\newblock ISSN 01466410.
\newblock URL
  \url{https://www.sciencedirect.com/science/article/pii/S0146641019300298}.

\bibitem[Chadwick et~al.(2011)Chadwick, Herman, Obložinský, Dunn, Danon,
  Kahler, Smith, Pritychenko, Arbanas, Arcilla, Brewer, Brown, Capote, Carlson,
  Cho, Derrien, Guber, Hale, Hoblit, Holloway, Johnson, Kawano, Kiedrowski,
  Kim, Kunieda, Larson, Leal, Lestone, Little, McCutchan, MacFarlane, MacInnes,
  Mattoon, McKnight, Mughabghab, Nobre, Palmiotti, Palumbo, Pigni, Pronyaev,
  Sayer, Sonzogni, Summers, Talou, Thompson, Trkov, Vogt, van~der Marck,
  Wallner, White, Wiarda, and Young]{ENDF_BVII.1}
M.B. Chadwick, M.~Herman, P.~Obložinský, M.E. Dunn, Y.~Danon, A.C. Kahler,
  D.L. Smith, B.~Pritychenko, G.~Arbanas, R.~Arcilla, R.~Brewer, D.A. Brown,
  R.~Capote, A.D. Carlson, Y.S. Cho, H.~Derrien, K.~Guber, G.M. Hale,
  S.~Hoblit, S.~Holloway, T.D. Johnson, T.~Kawano, B.C. Kiedrowski, H.~Kim,
  S.~Kunieda, N.M. Larson, L.~Leal, J.P. Lestone, R.C. Little, E.A. McCutchan,
  R.E. MacFarlane, M.~MacInnes, C.M. Mattoon, R.D. McKnight, S.F. Mughabghab,
  G.P.A. Nobre, G.~Palmiotti, A.~Palumbo, M.T. Pigni, V.G. Pronyaev, R.O.
  Sayer, A.A. Sonzogni, N.C. Summers, P.~Talou, I.J. Thompson, A.~Trkov, R.L.
  Vogt, S.C. van~der Marck, A.~Wallner, M.C. White, D.~Wiarda, and P.G. Young.
\newblock Endf/b-vii.1 nuclear data for science and technology: Cross sections,
  covariances, fission product yields and decay data.
\newblock \emph{Nuclear Data Sheets}, 112\penalty0 (12):\penalty0 2887 -- 2996,
  2011.
\newblock ISSN 0090-3752.
\newblock \doi{https://doi.org/10.1016/j.nds.2011.11.002}.
\newblock URL
  \url{http://www.sciencedirect.com/science/article/pii/S009037521100113X}.
\newblock Special Issue on ENDF/B-VII.1 Library.

\bibitem[Chadwick(2006)]{Chadwick2011}
M.B. Chadwick.
\newblock Report of the nuclear physics and related computational science r\&d
  for advanced fuel cycles workshop, August 2006.

\bibitem[Goriely et~al.(2019)Goriely, Dimitriou, Wiedeking, Belgya, Firestone,
  Kopecky, Krti{\v{c}}ka, Plujko, Schwengner, Siem, Utsunomiya, Hilaire,
  P{\'{e}}ru, Cho, Filipescu, Iwamoto, Kawano, Varlamov, and Xu]{Goriely2019}
S~Goriely, P~Dimitriou, M~Wiedeking, T~Belgya, R~Firestone, J~Kopecky,
  M~Krti{\v{c}}ka, V~Plujko, R~Schwengner, S~Siem, H~Utsunomiya, S~Hilaire,
  S~P{\'{e}}ru, Y~S Cho, D~M Filipescu, N~Iwamoto, T~Kawano, V~Varlamov, and
  R~Xu.
\newblock {Reference database for photon strength functions}.
\newblock \emph{The European Physical Journal A}, 55\penalty0 (10):\penalty0
  172, oct 2019.
\newblock ISSN 1434-601X.
\newblock \doi{10.1140/epja/i2019-12840-1}.
\newblock URL \url{https://doi.org/10.1140/epja/i2019-12840-1}.

\bibitem[Schiller et~al.(2000)Schiller, Bergholt, Guttormsen, Melby, Rekstad,
  and Siem]{OsloMethodNIM}
A~Schiller, L~Bergholt, M~Guttormsen, E~Melby, J~Rekstad, and S~Siem.
\newblock Extraction of level density and $\gamma$ strength function from
  primary $\gamma$ spectra.
\newblock \emph{Nucl. Instrum. Methods Phys. Res., Sect. A}, 447\penalty0
  (3):\penalty0 498--511, jun 2000.
\newblock ISSN 01689002.
\newblock \doi{10.1016/S0168-9002(99)01187-0}.
\newblock URL \url{https://dx.doi.org/10.1016/S0168-9002(99)01187-0}.

\bibitem[Guttormsen et~al.(2017)Guttormsen, Goriely, Larsen, G\"orgen, Hagen,
  Renstr\o{}m, Siem, Syed, Tagliente, Toft, Utsunomiya, Voinov, and
  Wikan]{PhysRevC.96.024313}
M.~Guttormsen, S.~Goriely, A.~C. Larsen, A.~G\"orgen, T.~W. Hagen,
  T.~Renstr\o{}m, S.~Siem, N.~U.~H. Syed, G.~Tagliente, H.~K. Toft,
  H.~Utsunomiya, A.~V. Voinov, and K.~Wikan.
\newblock Quasicontinuum $\ensuremath{\gamma}$ decay of $^{91,92}\mathbf{Zr}$:
  Benchmarking indirect ($n,\ensuremath{\gamma}$) cross section measurements
  for the $s$ process.
\newblock \emph{Phys. Rev. C}, 96:\penalty0 024313, Aug 2017.
\newblock \doi{10.1103/PhysRevC.96.024313}.
\newblock URL \url{https://link.aps.org/doi/10.1103/PhysRevC.96.024313}.

\bibitem[Kheswa et~al.(2017)Kheswa, Wiedeking, Brown, Larsen, Goriely,
  Guttormsen, Bello~Garrote, Bernstein, Bleuel, Eriksen, Giacoppo, G\"orgen,
  Goldblum, Hagen, Koehler, Klintefjord, Malatji, Midtb\o{}, Nyhus, Papka,
  Renstr\o{}m, Rose, Sahin, Siem, and Tornyi]{PhysRevC.95.045805}
B.~V. Kheswa, M.~Wiedeking, J.~A. Brown, A.~C. Larsen, S.~Goriely,
  M.~Guttormsen, F.~L. Bello~Garrote, L.~A. Bernstein, D.~L. Bleuel, T.~K.
  Eriksen, F.~Giacoppo, A.~G\"orgen, B.~L. Goldblum, T.~W. Hagen, P.~E.
  Koehler, M.~Klintefjord, K.~L. Malatji, J.~E. Midtb\o{}, H.~T. Nyhus,
  P.~Papka, T.~Renstr\o{}m, S.~J. Rose, E.~Sahin, S.~Siem, and T.~G. Tornyi.
\newblock $^{137,138,139}\mathbf{La}(n,\ensuremath{\gamma})$.
\newblock \emph{Phys. Rev. C}, 95:\penalty0 045805, Apr 2017.
\newblock \doi{10.1103/PhysRevC.95.045805}.
\newblock URL \url{https://link.aps.org/doi/10.1103/PhysRevC.95.045805}.

\bibitem[Larsen et~al.(2016)Larsen, Guttormsen, Schwengner, Bleuel, Goriely,
  Harissopulos, Bello~Garrote, Byun, Eriksen, Giacoppo, G\"orgen, Hagen,
  Klintefjord, Renstr\o{}m, Rose, Sahin, Siem, Tornyi, Tveten, Voinov, and
  Wiedeking]{PhysRevC.93.045810}
A.~C. Larsen, M.~Guttormsen, R.~Schwengner, D.~L. Bleuel, S.~Goriely,
  S.~Harissopulos, F.~L. Bello~Garrote, Y.~Byun, T.~K. Eriksen, F.~Giacoppo,
  A.~G\"orgen, T.~W. Hagen, M.~Klintefjord, T.~Renstr\o{}m, S.~J. Rose,
  E.~Sahin, S.~Siem, T.~G. Tornyi, G.~M. Tveten, A.~V. Voinov, and
  M.~Wiedeking.
\newblock Experimentally constrained $(p,\ensuremath{\gamma})^{89}\mathrm{Y}$
  and $(n,\ensuremath{\gamma})^{89}\mathrm{Y}$ reaction rates relevant to
  $p$-process nucleosynthesis.
\newblock \emph{Phys. Rev. C}, 93:\penalty0 045810, Apr 2016.
\newblock \doi{10.1103/PhysRevC.93.045810}.
\newblock URL \url{http://link.aps.org/doi/10.1103/PhysRevC.93.045810}.

\bibitem[Spyrou et~al.(2014)Spyrou, Liddick, Larsen, Guttormsen, Cooper,
  Dombos, Morrissey, Naqvi, Perdikakis, Quinn, Renstr\o{}m, Rodriguez, Simon,
  Sumithrarachchi, and Zegers]{PhysRevLett.113.232502}
A.~Spyrou, S.~N. Liddick, A.~C. Larsen, M.~Guttormsen, K.~Cooper, A.~C. Dombos,
  D.~J. Morrissey, F.~Naqvi, G.~Perdikakis, S.~J. Quinn, T.~Renstr\o{}m, J.~A.
  Rodriguez, A.~Simon, C.~S. Sumithrarachchi, and R.~G.~T. Zegers.
\newblock Novel technique for constraining $r$-process ($n$,
  $\ensuremath{\gamma}$) reaction rates.
\newblock \emph{Phys. Rev. Lett.}, 113:\penalty0 232502, Dec 2014.
\newblock \doi{10.1103/PhysRevLett.113.232502}.
\newblock URL \url{http://link.aps.org/doi/10.1103/PhysRevLett.113.232502}.

\bibitem[Koning et~al.(2008)Koning, Hilaire, and Duijvestijn]{Koning2007}
A.~J. Koning, S.~Hilaire, and M.~C. Duijvestijn.
\newblock {Talys-1.0}.
\newblock In \emph{International Conference on Nuclear Data for Science and
  Technology 2007}, pages 2--5, Les Ulis, France, may 2008. EDP Sciences.
\newblock \doi{10.1051/ndata:07767}.
\newblock URL \url{http://nd2007.edpsciences.org/10.1051/ndata:07767}.

\bibitem[Ltd.(2017)]{micron}
Micron~Semiconductor Ltd.
\newblock Product cataloque.
\newblock Technical report, Micron Semiconductor Ltd., 2017.
\newblock URL
  \url{http://micronsemiconductor.co.uk/wp-content/uploads/2017/01/cat.pdf}.

\bibitem[Lipoglavšek et~al.(2006)Lipoglavšek, Likar, Vencelj, Vidmar, Bark,
  Gueorguieva, Komati, Lawrie, Maliage, Mullins, Murray, and
  Ramashidzha]{NIMA2007}
M.~Lipoglavšek, A.~Likar, M.~Vencelj, T.~Vidmar, R.A. Bark, E.~Gueorguieva,
  F.~Komati, J.J. Lawrie, S.M. Maliage, S.M. Mullins, S.H.T. Murray, and T.M.
  Ramashidzha.
\newblock Measuring high-energy $\gamma$-rays with ge clover detectors.
\newblock \emph{Nuclear Instruments and Methods in Physics Research Section A:
  Accelerators, Spectrometers, Detectors and Associated Equipment},
  557\penalty0 (2):\penalty0 523 -- 527, 2006.
\newblock ISSN 0168-9002.
\newblock \doi{10.1016/j.nima.2005.11.067}.
\newblock URL
  \url{http://www.sciencedirect.com/science/article/pii/S0168900205021935}.

\bibitem[Tarasov and Bazin(2008)]{LisePP}
O.~B. Tarasov and D.~Bazin.
\newblock {LISE++: Radioactive beam production with in-flight separators}.
\newblock \emph{Nuclear Instruments and Methods in Physics Research, Section B:
  Beam Interactions with Materials and Atoms}, 266\penalty0 (19-20):\penalty0
  4657--4664, 2008.
\newblock ISSN 0168583X.
\newblock \doi{10.1016/j.nimb.2008.05.110}.
\newblock URL \url{http://lise.nscl.msu.edu}.

\bibitem[Guttormsen et~al.(1996)Guttormsen, Tveter, Bergholt, Ingebretsen, and
  Rekstad]{UnfoldingNIM}
M~Guttormsen, T.S Tveter, L~Bergholt, F~Ingebretsen, and J~Rekstad.
\newblock The unfolding of continuum $\gamma$-ray spectra.
\newblock \emph{Nucl. Instrum. Methods Phys. Res., Sect. A}, 374\penalty0
  (3):\penalty0 371--376, jun 1996.
\newblock ISSN 01689002.
\newblock \doi{10.1016/0168-9002(96)00197-0}.
\newblock URL \url{https://dx.doi.org/10.1016/0168-9002(96)00197-0}.

\bibitem[Agostinelli et~al.(2003)Agostinelli, Allison, Amako, Apostolakis,
  Araujo, Arce, Asai, Axen, Banerjee, Barrand, Behner, Bellagamba, Boudreau,
  Broglia, Brunengo, Burkhardt, Chauvie, Chuma, Chytracek, Cooperman, Cosmo,
  Degtyarenko, Dell'Acqua, Depaola, Dietrich, Enami, Feliciello, Ferguson,
  Fesefeldt, Folger, Foppiano, Forti, Garelli, Giani, Giannitrapani, Gibin,
  Cadenas, González, Abril, Greeniaus, Greiner, Grichine, Grossheim, Guatelli,
  Gumplinger, Hamatsu, Hashimoto, Hasui, Heikkinen, Howard, Ivanchenko,
  Johnson, Jones, Kallenbach, Kanaya, Kawabata, Kawabata, Kawaguti, Kelner,
  Kent, Kimura, Kodama, Kokoulin, Kossov, Kurashige, Lamanna, Lampén, Lara,
  Lefebure, Lei, Liendl, Lockman, Longo, Magni, Maire, Medernach, Minamimoto,
  de~Freitas, Morita, Murakami, Nagamatu, Nartallo, Nieminen, Nishimura,
  Ohtsubo, Okamura, O'Neale, Oohata, Paech, Perl, Pfeiffer, Pia, Ranjard,
  Rybin, Sadilov, Salvo, Santin, Sasaki, Savvas, Sawada, Scherer, Sei,
  Sirotenko, Smith, Starkov, Stoecker, Sulkimo, Takahata, Tanaka, Tcherniaev,
  Tehrani, Tropeano, Truscott, Uno, Urban, Urban, Verderi, Walkden, Wander,
  Weber, Wellisch, Wenaus, Williams, Wright, Yamada, Yoshida, and
  Zschiesche]{AGOSTINELLI2003250}
S.~Agostinelli, J.~Allison, K.~Amako, J.~Apostolakis, H.~Araujo, P.~Arce,
  M.~Asai, D.~Axen, S.~Banerjee, G.~Barrand, F.~Behner, L.~Bellagamba,
  J.~Boudreau, L.~Broglia, A.~Brunengo, H.~Burkhardt, S.~Chauvie, J.~Chuma,
  R.~Chytracek, G.~Cooperman, G.~Cosmo, P.~Degtyarenko, A.~Dell'Acqua,
  G.~Depaola, D.~Dietrich, R.~Enami, A.~Feliciello, C.~Ferguson, H.~Fesefeldt,
  G.~Folger, F.~Foppiano, A.~Forti, S.~Garelli, S.~Giani, R.~Giannitrapani,
  D.~Gibin, J.J.~Gómez Cadenas, I.~González, G.~Gracia Abril, G.~Greeniaus,
  W.~Greiner, V.~Grichine, A.~Grossheim, S.~Guatelli, P.~Gumplinger,
  R.~Hamatsu, K.~Hashimoto, H.~Hasui, A.~Heikkinen, A.~Howard, V.~Ivanchenko,
  A.~Johnson, F.W. Jones, J.~Kallenbach, N.~Kanaya, M.~Kawabata, Y.~Kawabata,
  M.~Kawaguti, S.~Kelner, P.~Kent, A.~Kimura, T.~Kodama, R.~Kokoulin,
  M.~Kossov, H.~Kurashige, E.~Lamanna, T.~Lampén, V.~Lara, V.~Lefebure,
  F.~Lei, M.~Liendl, W.~Lockman, F.~Longo, S.~Magni, M.~Maire, E.~Medernach,
  K.~Minamimoto, P.~Mora de~Freitas, Y.~Morita, K.~Murakami, M.~Nagamatu,
  R.~Nartallo, P.~Nieminen, T.~Nishimura, K.~Ohtsubo, M.~Okamura, S.~O'Neale,
  Y.~Oohata, K.~Paech, J.~Perl, A.~Pfeiffer, M.G. Pia, F.~Ranjard, A.~Rybin,
  S.~Sadilov, E.~Di Salvo, G.~Santin, T.~Sasaki, N.~Savvas, Y.~Sawada,
  S.~Scherer, S.~Sei, V.~Sirotenko, D.~Smith, N.~Starkov, H.~Stoecker,
  J.~Sulkimo, M.~Takahata, S.~Tanaka, E.~Tcherniaev, E.~Safai Tehrani,
  M.~Tropeano, P.~Truscott, H.~Uno, L.~Urban, P.~Urban, M.~Verderi, A.~Walkden,
  W.~Wander, H.~Weber, J.P. Wellisch, T.~Wenaus, D.C. Williams, D.~Wright,
  T.~Yamada, H.~Yoshida, and D.~Zschiesche.
\newblock Geant4—a simulation toolkit.
\newblock \emph{Nuclear Instruments and Methods in Physics Research Section A:
  Accelerators, Spectrometers, Detectors and Associated Equipment},
  506\penalty0 (3):\penalty0 250 -- 303, 2003.
\newblock ISSN 0168-9002.
\newblock \doi{10.1016/S0168-9002(03)01368-8}.
\newblock URL
  \url{http://www.sciencedirect.com/science/article/pii/S0168900203013688}.

\bibitem[Guttormsen et~al.(1987)Guttormsen, Ramsøy, and
  Rekstad]{FirstGenerationNIM}
M.~Guttormsen, T.~Ramsøy, and J.~Rekstad.
\newblock The first generation of $\gamma$-rays from hot nuclei.
\newblock \emph{Nucl. Instum. Methods Phys. Res., Sect. A}, 255\penalty0
  (3):\penalty0 518--523, apr 1987.
\newblock ISSN 01689002.
\newblock \doi{10.1016/0168-9002(87)91221-6}.
\newblock URL \url{https://dx.doi.org/10.1016/0168-9002(87)91221-6}.

\bibitem[Johnson and Kulp(2015)]{JOHNSON20151}
T.D. Johnson and W.D. Kulp.
\newblock Nuclear data sheets for a = 87.
\newblock \emph{Nuclear Data Sheets}, 129:\penalty0 1 -- 190, 2015.
\newblock ISSN 0090-3752.
\newblock \doi{10.1016/j.nds.2015.09.001}.
\newblock URL
  \url{http://www.sciencedirect.com/science/article/pii/S0090375215000460}.

\bibitem[Carlton et~al.(1988)Carlton, Winters, Johnson, Hill, and
  Harvey]{PhysRevC.38.1605}
R.~F. Carlton, R.~R. Winters, C.~H. Johnson, N.~W. Hill, and J.~A. Harvey.
\newblock Total cross section and resonance spectroscopy for n${+}^{86}$kr.
\newblock \emph{Phys. Rev. C}, 38:\penalty0 1605--1618, Oct 1988.
\newblock \doi{10.1103/PhysRevC.38.1605}.
\newblock URL \url{https://link.aps.org/doi/10.1103/PhysRevC.38.1605}.

\bibitem[Ericson and Strutinski(1958)]{Ericson1958}
T.~Ericson and V.~Strutinski.
\newblock {On angular distributions in compound nucleus processes}.
\newblock \emph{Nuclear Physics}, 8\penalty0 (C):\penalty0 284--293, sep 1958.
\newblock ISSN 00295582.
\newblock \doi{10.1016/0029-5582(58)90156-1}.
\newblock URL
  \url{https://www.sciencedirect.com/science/article/pii/0029558258901561}.

\bibitem[Egidy and Bucurescu(2005)]{PhysRevC.72.044311}
Till~von Egidy and Dorel Bucurescu.
\newblock Systematics of nuclear level density parameters.
\newblock \emph{Phys. Rev. C}, 72:\penalty0 044311, Oct 2005.
\newblock \doi{10.1103/PhysRevC.72.044311}.
\newblock URL \url{http://link.aps.org/doi/10.1103/PhysRevC.72.044311}.

\bibitem[Glibert and Cameron(1965)]{CanJPhys43.1446}
A.~Glibert and A.~G.~W. Cameron.
\newblock A composite nuclear-level density formula with shell corrections.
\newblock \emph{Can. J. Phys.}, 43:\penalty0 1446--1496, 1965.
\newblock \doi{10.1139/p65-139}.
\newblock URL \url{http://dx.doi.org/10.1139/p65-139}.

\bibitem[von Egidy and Bucurescu(2009)]{PhysRevC.80.054310}
T.~von Egidy and D.~Bucurescu.
\newblock Experimental energy-dependent nuclear spin distributions.
\newblock \emph{Phys. Rev. C}, 80:\penalty0 054310, Nov 2009.
\newblock \doi{10.1103/PhysRevC.80.054310}.
\newblock URL \url{https://link.aps.org/doi/10.1103/PhysRevC.80.054310}.

\bibitem[Goriely et~al.(2008)Goriely, Hilaire, and Koning]{PhysRevC.78.064307}
S.~Goriely, S.~Hilaire, and A.~J. Koning.
\newblock Improved microscopic nuclear level densities within the
  hartree-fock-bogoliubov plus combinatorial method.
\newblock \emph{Phys. Rev. C}, 78:\penalty0 064307, Dec 2008.
\newblock \doi{10.1103/PhysRevC.78.064307}.
\newblock URL \url{https://link.aps.org/doi/10.1103/PhysRevC.78.064307}.

\bibitem[Ericson(1960)]{doi:10.1080/00018736000101239}
Torleif Ericson.
\newblock {The statistical model and nuclear level densities}.
\newblock \emph{Advances in Physics}, 9\penalty0 (36):\penalty0 425--511, 1960.
\newblock \doi{10.1080/00018736000101239}.
\newblock URL \url{https://doi.org/10.1080/00018736000101239}.

\bibitem[Erba et~al.(1961)Erba, Facchini, and Menichella]{Erba1961}
E.~Erba, U.~Facchini, and E.S. Menichella.
\newblock {Statistical Emission in Nuclear Reactions and nuclear level
  density}.
\newblock \emph{Il Nuovo Cimento}, 22\penalty0 (6):\penalty0 1237--1260, 1961.
\newblock \doi{10.1007/BF02786895}.

\bibitem[Larsen et~al.(2011)Larsen, Guttormsen, Krti\ifmmode~\check{c}\else
  \v{c}\fi{}ka, B\ifmmode~\check{e}\else \v{e}\fi{}t\'ak, B\"urger, G\"orgen,
  Nyhus, Rekstad, Schiller, Siem, Toft, Tveten, Voinov, and
  Wikan]{PhysRevC.83.034315}
A.~C. Larsen, M.~Guttormsen, M.~Krti\ifmmode~\check{c}\else \v{c}\fi{}ka,
  E.~B\ifmmode~\check{e}\else \v{e}\fi{}t\'ak, A.~B\"urger, A.~G\"orgen, H.~T.
  Nyhus, J.~Rekstad, A.~Schiller, S.~Siem, H.~K. Toft, G.~M. Tveten, A.~V.
  Voinov, and K.~Wikan.
\newblock Analysis of possible systematic errors in the oslo method.
\newblock \emph{Phys. Rev. C}, 83:\penalty0 034315, Mar 2011.
\newblock \doi{10.1103/PhysRevC.83.034315}.
\newblock URL \url{https://link.aps.org/doi/10.1103/PhysRevC.83.034315}.

\bibitem[Raman et~al.(1983)Raman, Fogelberg, Harvey, Macklin, Stelson,
  Schr\"oder, and Kratz]{PhysRevC.28.602}
S.~Raman, B.~Fogelberg, J.~A. Harvey, R.~L. Macklin, P.~H. Stelson,
  A.~Schr\"oder, and K.~L. Kratz.
\newblock Overlapping $\ensuremath{\beta}$ decay and resonance neutron
  spectroscopy of levels in $^{87}\mathrm{Kr}$.
\newblock \emph{Phys. Rev. C}, 28:\penalty0 602--622, Aug 1983.
\newblock \doi{10.1103/PhysRevC.28.602}.
\newblock URL \url{https://link.aps.org/doi/10.1103/PhysRevC.28.602}.

\bibitem[Voinov et~al.(2001)Voinov, Guttormsen, Melby, Rekstad, Schiller, and
  Siem]{PhysRevC.63.044313}
A.~Voinov, M.~Guttormsen, E.~Melby, J.~Rekstad, A.~Schiller, and S.~Siem.
\newblock $\ensuremath{\gamma}$.
\newblock \emph{Phys. Rev. C}, 63:\penalty0 044313, Mar 2001.
\newblock \doi{10.1103/PhysRevC.63.044313}.
\newblock URL \url{http://link.aps.org/doi/10.1103/PhysRevC.63.044313}.

\bibitem[Schwengner et~al.(2013{\natexlab{a}})Schwengner, Massarczyk, Rusev,
  Tsoneva, Bemmerer, Beyer, Hannaske, Junghans, Kelley, Kwan, Lenske, Marta,
  Raut, Schilling, Tonchev, Tornow, and Wagner]{PhysRevC.87.024306}
R.~Schwengner, R.~Massarczyk, G.~Rusev, N.~Tsoneva, D.~Bemmerer, R.~Beyer,
  R.~Hannaske, A.~R. Junghans, J.~H. Kelley, E.~Kwan, H.~Lenske, M.~Marta,
  R.~Raut, K.~D. Schilling, A.~Tonchev, W.~Tornow, and A.~Wagner.
\newblock Pygmy dipole strength in ${}^{86}$kr and systematics of $n=50$
  isotones.
\newblock \emph{Phys. Rev. C}, 87:\penalty0 024306, Feb 2013{\natexlab{a}}.
\newblock \doi{10.1103/PhysRevC.87.024306}.
\newblock URL \url{http://link.aps.org/doi/10.1103/PhysRevC.87.024306}.

\bibitem[Raut et~al.(2013)Raut, Tonchev, Rusev, Tornow, Iliadis, Lugaro,
  Buntain, Goriely, Kelley, Schwengner, Banu, and
  Tsoneva]{PhysRevLett.111.112501}
R.~Raut, A.~P. Tonchev, G.~Rusev, W.~Tornow, C.~Iliadis, M.~Lugaro, J.~Buntain,
  S.~Goriely, J.~H. Kelley, R.~Schwengner, A.~Banu, and N.~Tsoneva.
\newblock Cross-section measurements of the
  $^{86}\mathrm{Kr}(\ensuremath{\gamma},n)$ reaction to probe the $s$-process
  branching at $^{85}\mathrm{Kr}$.
\newblock \emph{Phys. Rev. Lett.}, 111:\penalty0 112501, Sep 2013.
\newblock \doi{10.1103/PhysRevLett.111.112501}.
\newblock URL \url{http://link.aps.org/doi/10.1103/PhysRevLett.111.112501}.

\bibitem[Goriely et~al.(2004)Goriely, Khan, and Samyn]{GORIELY2004331}
S~Goriely, E~Khan, and M~Samyn.
\newblock Microscopic hfb + qrpa predictions of dipole strength for
  astrophysics applications.
\newblock \emph{Nuclear Physics A}, 739\penalty0 (3):\penalty0 331 -- 352,
  2004.
\newblock ISSN 0375-9474.
\newblock \doi{https://doi.org/10.1016/j.nuclphysa.2004.04.105}.
\newblock URL
  \url{http://www.sciencedirect.com/science/article/pii/S0375947404006578}.

\bibitem[Renstr\o{}m et~al.(2016)Renstr\o{}m, Nyhus, Utsunomiya, Schwengner,
  Goriely, Larsen, Filipescu, Gheorghe, Bernstein, Bleuel, Glodariu, G\"orgen,
  Guttormsen, Hagen, Kheswa, Lui, Negi, Ruud, Shima, Siem, Takahisa, Tesileanu,
  Tornyi, Tveten, and Wiedeking]{PhysRevC.93.064302}
T.~Renstr\o{}m, H.-T. Nyhus, H.~Utsunomiya, R.~Schwengner, S.~Goriely, A.~C.
  Larsen, D.~M. Filipescu, I.~Gheorghe, L.~A. Bernstein, D.~L. Bleuel,
  T.~Glodariu, A.~G\"orgen, M.~Guttormsen, T.~W. Hagen, B.~V. Kheswa, Y.-W.
  Lui, D.~Negi, I.~E. Ruud, T.~Shima, S.~Siem, K.~Takahisa, O.~Tesileanu, T.~G.
  Tornyi, G.~M. Tveten, and M.~Wiedeking.
\newblock Low-energy enhancement in the $\ensuremath{\gamma}$-ray strength
  functions of $^{73,74}\mathrm{Ge}$.
\newblock \emph{Phys. Rev. C}, 93:\penalty0 064302, Jun 2016.
\newblock \doi{10.1103/PhysRevC.93.064302}.
\newblock URL \url{http://link.aps.org/doi/10.1103/PhysRevC.93.064302}.

\bibitem[Guttormsen et~al.(2005)Guttormsen, Chankova, Agvaanluvsan, Algin,
  Bernstein, Ingebretsen, L\"onnroth, Messelt, Mitchell, Rekstad, Schiller,
  Siem, Sunde, Voinov, and \O{}deg\aa{}rd]{PhysRevC.71.044307}
M.~Guttormsen, R.~Chankova, U.~Agvaanluvsan, E.~Algin, L.~A. Bernstein,
  F.~Ingebretsen, T.~L\"onnroth, S.~Messelt, G.~E. Mitchell, J.~Rekstad,
  A.~Schiller, S.~Siem, A.~C. Sunde, A.~Voinov, and S.~\O{}deg\aa{}rd.
\newblock Radiative strength functions in $^{93\ensuremath{-}98}\mathrm{Mo}$.
\newblock \emph{Phys. Rev. C}, 71:\penalty0 044307, Apr 2005.
\newblock \doi{10.1103/PhysRevC.71.044307}.
\newblock URL \url{http://link.aps.org/doi/10.1103/PhysRevC.71.044307}.

\bibitem[Larsen et~al.(2006)Larsen, Chankova, Guttormsen, Ingebretsen, Messelt,
  Rekstad, Siem, Syed, \O{}deg\aa{}rd, L\"onnroth, Schiller, and
  Voinov]{PhysRevC.73.064301}
A.~C. Larsen, R.~Chankova, M.~Guttormsen, F.~Ingebretsen, S.~Messelt,
  J.~Rekstad, S.~Siem, N.~U.~H. Syed, S.~W. \O{}deg\aa{}rd, T.~L\"onnroth,
  A.~Schiller, and A.~Voinov.
\newblock Microcanonical entropies and radiative strength functions of
  $^{50,51}\mathrm{V}$.
\newblock \emph{Phys. Rev. C}, 73:\penalty0 064301, Jun 2006.
\newblock \doi{10.1103/PhysRevC.73.064301}.
\newblock URL \url{http://link.aps.org/doi/10.1103/PhysRevC.73.064301}.

\bibitem[Syed et~al.(2009)Syed, Larsen, B\"urger, Guttormsen, Harissopulos,
  Kmiecik, Konstantinopoulos, Krti\ifmmode~\check{c}\else \v{c}\fi{}ka,
  Lagoyannis, L\"onnroth, Mazurek, Norby, Nyhus, Perdikakis, Siem, and
  Spyrou]{PhysRevC.80.044309}
N.~U.~H. Syed, A.~C. Larsen, A.~B\"urger, M.~Guttormsen, S.~Harissopulos,
  M.~Kmiecik, T.~Konstantinopoulos, M.~Krti\ifmmode~\check{c}\else
  \v{c}\fi{}ka, A.~Lagoyannis, T.~L\"onnroth, K.~Mazurek, M.~Norby, H.~T.
  Nyhus, G.~Perdikakis, S.~Siem, and A.~Spyrou.
\newblock Extraction of thermal and electromagnetic properties in
  $^{45}\mathrm{Ti}$.
\newblock \emph{Phys. Rev. C}, 80:\penalty0 044309, Oct 2009.
\newblock \doi{10.1103/PhysRevC.80.044309}.
\newblock URL \url{http://link.aps.org/doi/10.1103/PhysRevC.80.044309}.

\bibitem[Larsen et~al.(2013{\natexlab{a}})Larsen, Ruud, B\"urger, Goriely,
  Guttormsen, G\"orgen, Hagen, Harissopulos, Nyhus, Renstr\o{}m, Schiller,
  Siem, Tveten, Voinov, and Wiedeking]{PhysRevC.87.014319}
A.~C. Larsen, I.~E. Ruud, A.~B\"urger, S.~Goriely, M.~Guttormsen, A.~G\"orgen,
  T.~W. Hagen, S.~Harissopulos, H.~T. Nyhus, T.~Renstr\o{}m, A.~Schiller,
  S.~Siem, G.~M. Tveten, A.~Voinov, and M.~Wiedeking.
\newblock Transitional $\ensuremath{\gamma}$ strength in cd isotopes.
\newblock \emph{Phys. Rev. C}, 87:\penalty0 014319, Jan 2013{\natexlab{a}}.
\newblock \doi{10.1103/PhysRevC.87.014319}.
\newblock URL \url{http://link.aps.org/doi/10.1103/PhysRevC.87.014319}.

\bibitem[Simon et~al.(2016)Simon, Guttormsen, Larsen, Beausang, Humby, Burke,
  Casperson, Hughes, Ross, Allmond, Chyzh, Dag, Koglin, McCleskey, McCleskey,
  Ota, and Saastamoinen]{PhysRevC.93.034303}
A.~Simon, M.~Guttormsen, A.~C. Larsen, C.~W. Beausang, P.~Humby, J.~T. Burke,
  R.~J. Casperson, R.~O. Hughes, T.~J. Ross, J.~M. Allmond, R.~Chyzh, M.~Dag,
  J.~Koglin, E.~McCleskey, M.~McCleskey, S.~Ota, and A.~Saastamoinen.
\newblock First observation of low-energy $\ensuremath{\gamma}$-ray enhancement
  in the rare-earth region.
\newblock \emph{Phys. Rev. C}, 93:\penalty0 034303, Mar 2016.
\newblock \doi{10.1103/PhysRevC.93.034303}.
\newblock URL \url{http://link.aps.org/doi/10.1103/PhysRevC.93.034303}.

\bibitem[Kheswa et~al.(2015)Kheswa, Wiedeking, Giacoppo, Goriely, Guttormsen,
  Larsen, Garrote, Eriksen, Görgen, Hagen, Koehler, Klintefjord, Nyhus, Papka,
  Renstrøm, Rose, Sahin, Siem, and Tornyi]{Kheswa2015268}
B.~V. Kheswa, M.~Wiedeking, F.~Giacoppo, S.~Goriely, M.~Guttormsen, A.C.
  Larsen, F.L.~Bello Garrote, T.K. Eriksen, A.~Görgen, T.W. Hagen, P.E.
  Koehler, M.~Klintefjord, H.T. Nyhus, P.~Papka, T.~Renstrøm, S.~Rose,
  E.~Sahin, S.~Siem, and T.~Tornyi.
\newblock Galactic production of $^{138}\mathrm{La}$: Impact of
  $^{138,139}\mathrm{La}$ statistical properties.
\newblock \emph{Phys. Lett. B}, 744:\penalty0 268 -- 272, 2015.
\newblock ISSN 0370-2693.
\newblock \doi{10.1016/j.physletb.2015.03.065}.
\newblock URL
  \url{http://www.sciencedirect.com/science/article/pii/S0370269315002403}.

\bibitem[Wiedeking et~al.(2012)Wiedeking, Bernstein,
  Krti\ifmmode~\check{c}\else \v{c}\fi{}ka, Bleuel, Allmond, Basunia, Burke,
  Fallon, Firestone, Goldblum, Hatarik, Lake, Lee, Lesher, Paschalis, Petri,
  Phair, and Scielzo]{PhysRevLett.108.162503}
M.~Wiedeking, L.~A. Bernstein, M.~Krti\ifmmode~\check{c}\else \v{c}\fi{}ka,
  D.~L. Bleuel, J.~M. Allmond, M.~S. Basunia, J.~T. Burke, P.~Fallon, R.~B.
  Firestone, B.~L. Goldblum, R.~Hatarik, P.~T. Lake, I-Y. Lee, S.~R. Lesher,
  S.~Paschalis, M.~Petri, L.~Phair, and N.~D. Scielzo.
\newblock Low-energy enhancement in the photon strength of $^{95}\mathrm{Mo}$.
\newblock \emph{Phys. Rev. Lett.}, 108:\penalty0 162503, Apr 2012.
\newblock \doi{10.1103/PhysRevLett.108.162503}.
\newblock URL \url{http://link.aps.org/doi/10.1103/PhysRevLett.108.162503}.

\bibitem[Larsen et~al.(2013{\natexlab{b}})Larsen, Blasi, Bracco, Camera,
  Eriksen, G\"orgen, Guttormsen, Hagen, Leoni, Million, Nyhus, Renstr\o{}m,
  Rose, Ruud, Siem, Tornyi, Tveten, Voinov, and
  Wiedeking]{PhysRevLett.111.242504}
A.~C. Larsen, N.~Blasi, A.~Bracco, F.~Camera, T.~K. Eriksen, A.~G\"orgen,
  M.~Guttormsen, T.~W. Hagen, S.~Leoni, B.~Million, H.~T. Nyhus,
  T.~Renstr\o{}m, S.~J. Rose, I.~E. Ruud, S.~Siem, T.~Tornyi, G.~M. Tveten,
  A.~V. Voinov, and M.~Wiedeking.
\newblock Evidence for the dipole nature of the low-energy
  $\ensuremath{\gamma}$ enhancement in $^{56}\mathrm{Fe}$.
\newblock \emph{Phys. Rev. Lett.}, 111:\penalty0 242504, Dec
  2013{\natexlab{b}}.
\newblock \doi{10.1103/PhysRevLett.111.242504}.
\newblock URL \url{http://link.aps.org/doi/10.1103/PhysRevLett.111.242504}.

\bibitem[Larsen et~al.(2017)Larsen, Guttormsen, Blasi, Bracco, Camera, Campo,
  Eriksen, Görgen, Hagen, Ingeberg, Kheswa, Leoni, Midtbø, Million, Nyhus,
  Renstrøm, Rose, Ruud, Siem, Tornyi, Tveten, Voinov, Wiedeking, and
  Zeiser]{Lar2017}
A~C Larsen, M~Guttormsen, N~Blasi, A~Bracco, F~Camera, L~Crespo Campo, T~K
  Eriksen, A~Görgen, T~W Hagen, V~W Ingeberg, B~V Kheswa, S~Leoni, J~E
  Midtbø, B~Million, H~T Nyhus, T~Renstrøm, S~J Rose, I~E Ruud, S~Siem, T~G
  Tornyi, G~M Tveten, A~V Voinov, M~Wiedeking, and F~Zeiser.
\newblock Low-energy enhancement and fluctuations of $\gamma$-ray strength
  functions in 56,57 fe: test of the brink–axel hypothesis.
\newblock \emph{Jour. Phys. G Nucl. and Part. Phys.}, 44\penalty0 (6):\penalty0
  064005, 2017.
\newblock URL \url{http://stacks.iop.org/0954-3899/44/i=6/a=064005}.

\bibitem[Jones et~al.(2018)Jones, Macchiavelli, Wiedeking, Bernstein, Crawford,
  Campbell, Clark, Cromaz, Fallon, Lee, Salathe, Wiens, Ayangeakaa, Bleuel,
  Bottoni, Carpenter, Davids, Elson, G\"orgen, Guttormsen, Janssens, Kinnison,
  Kirsch, Larsen, Lauritsen, Reviol, Sarantites, Siem, Voinov, and
  Zhu]{PhysRevC.97.024327}
M.~D. Jones, A.~O. Macchiavelli, M.~Wiedeking, L.~A. Bernstein, H.~L. Crawford,
  C.~M. Campbell, R.~M. Clark, M.~Cromaz, P.~Fallon, I.~Y. Lee, M.~Salathe,
  A.~Wiens, A.~D. Ayangeakaa, D.~L. Bleuel, S.~Bottoni, M.~P. Carpenter, H.~M.
  Davids, J.~Elson, A.~G\"orgen, M.~Guttormsen, R.~V.~F. Janssens, J.~E.
  Kinnison, L.~Kirsch, A.~C. Larsen, T.~Lauritsen, W.~Reviol, D.~G. Sarantites,
  S.~Siem, A.~V. Voinov, and S.~Zhu.
\newblock Examination of the low-energy enhancement of the
  $\ensuremath{\gamma}$-ray strength function of $^{56}\mathrm{Fe}$.
\newblock \emph{Phys. Rev. C}, 97:\penalty0 024327, Feb 2018.
\newblock \doi{10.1103/PhysRevC.97.024327}.
\newblock URL \url{https://link.aps.org/doi/10.1103/PhysRevC.97.024327}.

\bibitem[Larsen and Goriely(2010)]{PhysRevC.82.014318}
A.~C. Larsen and S.~Goriely.
\newblock Impact of a low-energy enhancement in the $\ensuremath{\gamma}$-ray
  strength function on the neutron-capture cross section.
\newblock \emph{Phys. Rev. C}, 82:\penalty0 014318, Jul 2010.
\newblock \doi{10.1103/PhysRevC.82.014318}.
\newblock URL \url{http://link.aps.org/doi/10.1103/PhysRevC.82.014318}.

\bibitem[Schwengner et~al.(2013{\natexlab{b}})Schwengner, Frauendorf, and
  Larsen]{PhysRevLett.111.232504}
R.~Schwengner, S.~Frauendorf, and A.~C. Larsen.
\newblock Low-energy enhancement of magnetic dipole radiation.
\newblock \emph{Phys. Rev. Lett.}, 111:\penalty0 232504, Dec
  2013{\natexlab{b}}.
\newblock \doi{10.1103/PhysRevLett.111.232504}.
\newblock URL \url{https://link.aps.org/doi/10.1103/PhysRevLett.111.232504}.

\bibitem[Czerwi\ifmmode~\acute{n}\else \'{n}\fi{}ski
  et~al.(2015)Czerwi\ifmmode~\acute{n}\else \'{n}\fi{}ski, Rz\k{a}ca-Urban,
  Urban, B\k{a}czyk, Sieja, Nyak\'o, Tim\'ar, Kuti, Tornyi, Atanasova, Blanc,
  Jentschel, Mutti, K\"oster, Soldner, de~France, Simpson, and
  Ur]{PhysRevC.92.014328}
M.~Czerwi\ifmmode~\acute{n}\else \'{n}\fi{}ski, T.~Rz\k{a}ca-Urban, W.~Urban,
  P.~B\k{a}czyk, K.~Sieja, B.~M. Nyak\'o, J.~Tim\'ar, I.~Kuti, T.~G. Tornyi,
  L.~Atanasova, A.~Blanc, M.~Jentschel, P.~Mutti, U.~K\"oster, T.~Soldner,
  G.~de~France, G.~S. Simpson, and C.~A. Ur.
\newblock Neutron-proton multiplets in the nucleus $^{88}\mathrm{Br}$.
\newblock \emph{Phys. Rev. C}, 92:\penalty0 014328, Jul 2015.
\newblock \doi{10.1103/PhysRevC.92.014328}.
\newblock URL \url{https://link.aps.org/doi/10.1103/PhysRevC.92.014328}.

\bibitem[Litzinger et~al.(2015)Litzinger, Blazhev, Dewald, Didierjean,
  Duch\^ene, Fransen, Lozeva, Sieja, Verney, de~Angelis, Bazzacco, Birkenbach,
  Bottoni, Bracco, Braunroth, Cederwall, Corradi, Crespi, D\'esesquelles,
  Eberth, Ellinger, Farnea, Fioretto, Gernh\"auser, Goasduff, G\"orgen,
  Gottardo, Grebosz, Hackstein, Hess, Ibrahim, Jolie, Jungclaus, Kolos, Korten,
  Leoni, Lunardi, Maj, Menegazzo, Mengoni, Michelagnoli, Mijatovic, Million,
  M\"oller, Modamio, Montagnoli, Montanari, Morales, Napoli, Niikura,
  Pollarolo, Pullia, Quintana, Recchia, Reiter, Rosso, Sahin, Salsac,
  Scarlassara, S\"oderstr\"om, Stefanini, Stezowski, Szilner, Theisen,
  Valiente~Dob\'on, Vandone, and Vogt]{PhysRevC.92.064322}
J.~Litzinger, A.~Blazhev, A.~Dewald, F.~Didierjean, G.~Duch\^ene, C.~Fransen,
  R.~Lozeva, K.~Sieja, D.~Verney, G.~de~Angelis, D.~Bazzacco, B.~Birkenbach,
  S.~Bottoni, A.~Bracco, T.~Braunroth, B.~Cederwall, L.~Corradi, F.~C.~L.
  Crespi, P.~D\'esesquelles, J.~Eberth, E.~Ellinger, E.~Farnea, E.~Fioretto,
  R.~Gernh\"auser, A.~Goasduff, A.~G\"orgen, A.~Gottardo, J.~Grebosz,
  M.~Hackstein, H.~Hess, F.~Ibrahim, J.~Jolie, A.~Jungclaus, K.~Kolos,
  W.~Korten, S.~Leoni, S.~Lunardi, A.~Maj, R.~Menegazzo, D.~Mengoni,
  C.~Michelagnoli, T.~Mijatovic, B.~Million, O.~M\"oller, V.~Modamio,
  G.~Montagnoli, D.~Montanari, A.~I. Morales, D.~R. Napoli, M.~Niikura,
  G.~Pollarolo, A.~Pullia, B.~Quintana, F.~Recchia, P.~Reiter, D.~Rosso,
  E.~Sahin, M.~D. Salsac, F.~Scarlassara, P.-A. S\"oderstr\"om, A.~M.
  Stefanini, O.~Stezowski, S.~Szilner, Ch. Theisen, J.~J. Valiente~Dob\'on,
  V.~Vandone, and A.~Vogt.
\newblock Transition probabilities in neutron-rich $^{84,86}\mathrm{Se}$.
\newblock \emph{Phys. Rev. C}, 92:\penalty0 064322, Dec 2015.
\newblock \doi{10.1103/PhysRevC.92.064322}.
\newblock URL \url{https://link.aps.org/doi/10.1103/PhysRevC.92.064322}.

\bibitem[Caurier et~al.(2005)Caurier, Mart\'{\i}nez-Pinedo, Nowacki, Poves, and
  Zuker]{RevModPhys.77.427}
E.~Caurier, G.~Mart\'{\i}nez-Pinedo, F.~Nowacki, A.~Poves, and A.~P. Zuker.
\newblock The shell model as a unified view of nuclear structure.
\newblock \emph{Rev. Mod. Phys.}, 77:\penalty0 427--488, Jun 2005.
\newblock \doi{10.1103/RevModPhys.77.427}.
\newblock URL \url{https://link.aps.org/doi/10.1103/RevModPhys.77.427}.

\bibitem[Sieja(2017)]{PhysRevLett.119.052502}
K.~Sieja.
\newblock Electric and magnetic dipole strength at low energy.
\newblock \emph{Phys. Rev. Lett.}, 119:\penalty0 052502, Jul 2017.
\newblock \doi{10.1103/PhysRevLett.119.052502}.
\newblock URL \url{https://link.aps.org/doi/10.1103/PhysRevLett.119.052502}.

\bibitem[Koning and Delaroche(2003)]{Koning2003}
A.J. Koning and J.P. Delaroche.
\newblock Local and global nucleon optical models from 1 kev to 200 mev.
\newblock \emph{Nucl. Phys. A}, 713\penalty0 (3):\penalty0 231 -- 310, 2003.
\newblock ISSN 0375-9474.
\newblock \doi{10.1016/S0375-9474(02)01321-0}.
\newblock URL
  \url{http://www.sciencedirect.com/science/article/pii/S0375947402013210}.

\bibitem[Hauser and Feshbach(1952)]{PhysRev.87.366}
Walter Hauser and Herman Feshbach.
\newblock The inelastic scattering of neutrons.
\newblock \emph{Phys. Rev.}, 87:\penalty0 366--373, Jul 1952.
\newblock \doi{10.1103/PhysRev.87.366}.
\newblock URL \url{http://link.aps.org/doi/10.1103/PhysRev.87.366}.

\bibitem[Bauge et~al.(2001)Bauge, Delaroche, and Girod]{PhysRevC.63.024607}
E.~Bauge, J.~P. Delaroche, and M.~Girod.
\newblock Lane-consistent, semimicroscopic nucleon-nucleus optical model.
\newblock \emph{Phys. Rev. C}, 63:\penalty0 024607, Jan 2001.
\newblock \doi{10.1103/PhysRevC.63.024607}.
\newblock URL \url{https://link.aps.org/doi/10.1103/PhysRevC.63.024607}.

\bibitem[Bhike et~al.(2015)Bhike, Rubino, Gooden, Krishichayan, and
  Tornow]{PhysRevC.92.014624}
Megha Bhike, E.~Rubino, M.~E. Gooden, Krishichayan, and W.~Tornow.
\newblock Measurements of the
  $^{86}\mathrm{Kr}(n,\ensuremath{\gamma})^{87}\mathrm{Kr}$ and
  $^{86}\mathrm{Kr}(n,2n)^{85}\mathrm{Kr}^{m}$ reaction cross sections below
  ${E}_{n}=15$ mev.
\newblock \emph{Phys. Rev. C}, 92:\penalty0 014624, Jul 2015.
\newblock \doi{10.1103/PhysRevC.92.014624}.
\newblock URL \url{http://link.aps.org/doi/10.1103/PhysRevC.92.014624}.

\bibitem[Walter et~al.(1986)Walter, Leugers, Käppeler, Bao, Reffo, and
  Fabbri]{NuclSciEng93.357}
G.~Walter, B.~Leugers, F.~Käppeler, Z.~Y. Bao, G.~Reffo, and F.~Fabbri.
\newblock Kilo-electron-volt neutron capture cross sections of the krypton
  isotopes.
\newblock \emph{Nuclear Science and Engineering}, 93:\penalty0 357--369, 1986.
\newblock \doi{10.13182/NSE86-A18471}.
\newblock URL \url{http://dx.doi.org/10.13182/NSE86-A18471}.

\bibitem[Beer et~al.(2002)Beer, Sedyshev, Rochow, Mohr, and
  Oberhummer]{Beer2002}
H.~Beer, P.~V. Sedyshev, W.~Rochow, P.~Mohr, and H.~Oberhummer.
\newblock {Neutron capture measurements of the noble gas isotopes22Ne, 40Ar and
  78,80,84,86Kr in the keV energy region}.
\newblock \emph{Nuclear Physics A}, 705\penalty0 (1-2):\penalty0 239--261, jul
  2002.
\newblock ISSN 03759474.
\newblock \doi{10.1016/S0375-9474(02)00645-0}.
\newblock URL
  \url{https://www.sciencedirect.com/science/article/abs/pii/S0375947402006450}.

\bibitem[Dillmann(2006)]{Dillmann2006}
I.~Dillmann.
\newblock Kadonis- the karlsruhe astrophysical database of nucleosynthesis in
  stars.
\newblock \emph{AIP Conf. Proc.}, 2006.
\newblock \doi{10.1063/1.2187846}.
\newblock URL \url{http://dx.doi.org/10.1063/1.2187846}.
\newblock online at http://www.kadonis.org.

\bibitem[Rauscher et~al.(1997)Rauscher, Thielemann, and
  Kratz]{PhysRevC.56.1613}
Thomas Rauscher, Friedrich~Karl Thielemann, and Karl~Ludwig Kratz.
\newblock {Nuclear level density and the determination of thermonuclear rates
  for astrophysics}.
\newblock \emph{Physical Review C - Nuclear Physics}, 56\penalty0 (3):\penalty0
  1613--1625, sep 1997.
\newblock ISSN 1089490X.
\newblock \doi{10.1103/PhysRevC.56.1613}.
\newblock URL \url{https://link.aps.org/doi/10.1103/PhysRevC.56.1613}.

\bibitem[Rochman et~al.(2013)Rochman, Koning, Kopecky, Sublet, Ribon, and
  Moxon]{Rochman2013}
D.~Rochman, A.~J. Koning, J.~Kopecky, J.~C. Sublet, P.~Ribon, and M.~Moxon.
\newblock {From average parameters to statistical resolved resonances}.
\newblock \emph{Annals of Nuclear Energy}, 51:\penalty0 60--68, jan 2013.
\newblock ISSN 03064549.
\newblock \doi{10.1016/j.anucene.2012.08.015}.
\newblock URL
  \url{https://www.sciencedirect.com/science/article/pii/S0306454912003350}.

\bibitem[Rochman et~al.(2017)Rochman, Goriely, Koning, and
  Ferroukhi]{Rochman2017}
D.~Rochman, S.~Goriely, A.J. Koning, and H.~Ferroukhi.
\newblock Radiative neutron capture: Hauser feshbach vs. statistical
  resonances.
\newblock \emph{Phys. Lett. B}, 764:\penalty0 109–113, Jan 2017.
\newblock \doi{10.1016/j.physletb.2016.11.018}.
\newblock URL \url{http://dx.doi.org/10.1016/j.physletb.2016.11.018}.

\bibitem[Krti\ifmmode~\check{c}\else \v{c}\fi{}ka
  et~al.(2016)Krti\ifmmode~\check{c}\else \v{c}\fi{}ka, Wiedeking, Be\ifmmode
  \check{c}\else \v{c}\fi{}v\'a\ifmmode~\check{r}\else \v{r}\fi{}, and
  Valenta]{PhysRevC.93.054311}
M.~Krti\ifmmode~\check{c}\else \v{c}\fi{}ka, M.~Wiedeking, F.~Be\ifmmode
  \check{c}\else \v{c}\fi{}v\'a\ifmmode~\check{r}\else \v{r}\fi{}, and
  S.~Valenta.
\newblock Consistency of photon strength function models with data from the
  $^{94}\mathrm{Mo}(d,\phantom{\rule{0.16em}{0ex}}p\ensuremath{\gamma}\ensuremath{\gamma})$
  reaction.
\newblock \emph{Phys. Rev. C}, 93:\penalty0 054311, May 2016.
\newblock \doi{10.1103/PhysRevC.93.054311}.
\newblock URL \url{https://link.aps.org/doi/10.1103/PhysRevC.93.054311}.

\end{thebibliography}
\end{document}